# Ferromagnetic and insulating behavior in both half magnetic levitation and non-levitation LK-99 like samples

Pinyuan Wang[1], Xiaoqi Liu[1], Jun Ge[1], Chengcheng Ji[1], Haoran Ji[1], Yanzhao Liu[1], Yiwen Ai[1], Gaoxing Ma[1], Shichao Qi[1], Jian Wang[1,2,3]*

[1]International Center for Quantum Materials, School of Physics, Peking University, Beijing 100871, China.
[2]Collaborative Innovation Center of Quantum Matter, Beijing 100871, China.
[3]Hefei National Laboratory, Hefei 230088, China.
*Corresponding author. Email: jianwangphysics@pku.edu.cn (J.W.).

**Finding materials exhibiting superconductivity at room temperature has long been one of the ultimate goals in physics and material science. Recently, room-temperature superconducting properties have been claimed in a copper substituted lead phosphate apatite ($Pb_{10-x}Cu_x(PO_4)_6O$, or called LK-99) [1-3]. Using a similar approach, we have prepared LK-99 like samples and confirmed the half-levitation behaviors in some small specimens under the influence of a magnet at room temperature. To examine the magnetic properties of our samples, we have performed systematic magnetization measurements on the as-grown LK-99 like samples, including the half-levitated and non-levitated samples. The magnetization measurements show the coexistence of soft-ferromagnetic and diamagnetic signals in both half-levitated and non-levitated samples. The electrical transport measurements on the as-grown LK-99 like samples including both half-levitated and non-levitated samples show an insulating behavior characterized by the increasing resistivity with the decreasing temperature.**

**Keywords**: $Pb_{10-x}Cu_x(PO_4)_6O$, LK-99, Superconductivity, Ferromagnetism, Insulating behavior, Half magnetic levitation

**Introduction**

Room-temperature superconductivity at ambient pressure is the long-sought goal of scientists for its possibility of revolution in scientific exploration, energy transportation and information technology[1-4]. Recently, a copper substituted lead phosphate apatite ($Pb_{10-x}Cu_x(PO_4)_6O$, or called LK-99) is claimed to exhibit possible room-temperature superconductivity [5-7] with the observation of an abrupt resistance drop at 105 °C, half magnetic levitation at room temperature and diamagnetic susceptibilities[5-7]. Subsequently, various theoretical [8-14] and experimental works [15-24] are inspired to study the properties of LK-99. Independent experiments have shown different magnetism and electrical transport properties. For example, Hou *et al.* detected nearly zero resistance below 100 K in their sample [17], which was subsequently explained by non-superconducting current percolation model by the same group [21]. Zhu *et al.* reported sharp resistance drop near 400 K, which was attributed to the structure transition of $Cu_2S$ impurity in their samples [20]. Guo *et al.* observed soft-ferromagnetism behavior in their half magnetic levitation samples[19].

In this work, polycrystalline LK-99 like samples were synthesized. Powder X-ray diffraction

(XRD) and energy-dispersive X-ray spectroscopy (EDS) measurements confirm that our samples consist of $Pb_{10-x}Cu_x(PO_4)_6O$ and $Cu_2S$, consistent with previous reports [7,19]. Remarkably, some small samples show half magnetic levitation at room temperature (Extended Video.1 for s2, Extended Video.2 for s3), similar with previous reports [5-7,19], while larger samples show no visible motion upon magnet. Spatial EDS mapping illustrates the distribution of $Pb_{10-x}Cu_x(PO_4)_6O$ and $Cu_2S$ is highly spatially inhomogeneous. The electrical transport measurements on the as-grown samples reveal a typical insulating behavior. To examine the magnetic properties of our sample, we have performed systematic magnetization measurements on the as-grown LK-99 like sample in a Magnetic Properties Measurement System (MPMS-3, Quantum Design). The magnetization measurements on the half-levitated and non-levitated samples show the coexistence of soft-ferromagnetic and diamagnetic signals. The saturated magnetization per unit mass at 300 K of half-levitated sample is found to be about twice larger than the non-levitated samples, which explains why some samples could levitate while some could not. No superconducting signal is detected in magnetism or electrical transport measurements.

**Materials and Methods**

Polycrystalline LK-99 like samples were grown by solid-state method. The growth process consists of three steps [7,17,19]. Firstly, to obtain precursor $Pb_3O_2SO_4$ [17], high-purity $Pb(SO)_4$ powder (Aladdin 99.99%) and PbO powder (Aladdin 99.999%) were mixed 1:1 stoichiometrically and then grinded thoroughly for 35 minutes. The mixture was transferred into an alumina crucible and then heated in the air for 24 hours with a heating rate of about 0.8°C/min. Pure white $Pb_3O_2SO_4$ power can be obtained after natural cooling to room temperature. Secondly, to obtain precursor $Cu_3P$, high-purity Cu powder (Aladdin 99.99%) and red phosphorus (Hhulhe 99.999%) were mixed with a molar ratio of 3:1 and grinded thoroughly for 35 minutes. The mixture was placed into an alumina crucible and then sealed in a quartz tube under vacuum condition ($<10^{-3}$ torr). After 48 hours of heating with a heating rate of about 0.6°C/min, black $Cu_3P$ was obtained. Finally, 1:1 stoichiometric mixture of the precursors $Pb_3O_2SO_4$ and $Cu_3P$ was grinded for 35 minutes in the air and then transferred into an alumina crucible. After sealing in a quartz tube under vacuum condition, the mixture was heated to 925°C in five hours and maintained for 10 hours. $Pb_{10-x}Cu_x(PO_4)_6O$ samples mixed with $Cu_2S$ were obtained in the quartz tube after cooling to room temperature in the furnace.

For the convenience of further selection and characterization of the as-grown samples, the sample was compressed into cylindrical thin slice at 30MPa. To find samples that can be levitated by an external magnetic field, the cylindrical thin slices were then grinded into smaller pieces (transverse size ranging from 0.2 mm to 10 mm). With the approaching of a $Nd_2Fe_{14}B$ magnet without direct contact with the sample container, some samples were half-levitated. The extended videos (Extended Video.1, Extended Video.2) show the motion of two half-levitation samples (s2 and s3) generated by a magnet, which is similar to previous report[7]. The samples without half-levitation (s1 and s4) were also picked up for further experiments.

The powder XRD measurements were performed in a Rigaku Mini-flux 600 X-ray diffractometer at room temperature and the EDS measurements were carried out in an FEI Helios NanoLab 600i DualBeam System.

To examine the magnetic properties of the LK-99 like samples, we performed systematic magnetization measurements in a Magnetic Properties Measurement System. In the magnetization measurements, samples were fixed on the quartz holder by GE Vanish. Our electrical transport measurements were conducted in a 16 T-Physical Property Measurement System (PPMS-16, Quantum Design).

**Experiments and Results**

As shown in Fig.1, the main XRD peaks correspond to $Pb_{10-x}Cu_x(PO_4)_6O$ and some other peaks represent the by-product, mainly including Cu and its sulphides [5-7, 19].

The optical image and EDS mappings of our non-levitated sample s1 are shown in Fig.2. Remarkably, the distribution of Cu and Pb (or P) is nearly spatially complementary. The EDS measurements show that the distribution of $Pb_{10-x}Cu_x(PO_4)_6O$ in our sample is inhomogeneous and influenced by Cu and its sulphides.

Figure 3 shows the electrical transport results and magnetic properties of non-levitated sample s1. As shown in the resistivity ($\rho$)-temperature ($T$) curve in Fig. 3a, the $\rho$-$T$ curve exhibits a typical insulating behavior, i.e. the resistivity exponentially increases with decreasing temperature (Fig.3a, b). The temperature-dependent resistivity is fitted with the band-gap model for insulators[25] ($\ln\rho \propto 1/T$, shown in Fig.3b). The fitted gap is ~228.49 meV. Figure 3c shows the magnetization versus magnetic field (*M-H*) curves at 2 K and 300 K. In the *M-H* measurements of each temperature, the external magnetic field is swept from 70 to -70 kOe, and subsequently from -70 to 70 kOe. When the magnetic field is below a certain field (~20 000 Oe at 2 K and ~2500 Oe at 300 K), the magnetization quickly increases with increasing the magnetic field. Further increasing the magnetic field, the magnetization tends to saturate, superposed with linear diamagnetic backgrounds at higher magnetic field. After subtracting the linear diamagnetic backgrounds, the saturated magnetization (~$0.6\times10^{-2}$ emu/g at 300 K and ~$3.9\times10^{-2}$ emu/g at 2 K) can be obtained (Fig.3d). More importantly, a hysteresis loop around zero magnetic field can be clearly observed in the *M-H* curve at 2 K, which slowly shrinks with increasing temperature (shown in the zoom-in *M-H* curves near zero magnetic field in Fig. 3e). The hysteresis loops indicate the existence of long-range ferromagnetic order in s1. It is worth noting that the hysteresis loop can be still detected at 400 K, suggesting a Curie temperature above 400 K. Fig. 3f shows zero-field-cooling (ZFC) and field-cooling (FC) *M-T* curves ranging from 2 K to 400 K under the applied external magnetic field of 100 Oe. Both the FC and ZFC curves show positive magnetic moments and an obvious branching can be observed below 400 K (Fig. 3f). These results indicate the existence of ferromagnetic order up to 400 K, consistent with our *M-H* results showing ferromagnetic hysteresis loops near zero magnetic field. In our results, the ZFC magnetization value remains positive in the whole temperature ranging from 400 K to 2 K and shows no abrupt change, suggesting that superconducting state is absent in this sample.

Figure 4 illustrates the optical image and EDS mappings of a smaller sample s2 showing half-levitation upon a magnet (as illustrated in Extended Video.1). Similar inhomogeneous distribution of Cu, S, O, Pb and P as s1 is observed. The distribution of Cu and Pb (or P) is also spatially complementary.

We then perform electrical transport and magnetization measurements on sample s2 in the same PPMS and MPMS instruments as s1. Figure 5a shows the $\rho$-$T$ curve of s2. Similar with s1, the resistivity increases with decreasing temperature, showing insulating behavior. The fitted gap with the band-gap model is ~207.52 meV(Fig.5b), close to the gap in s1(~228.49 meV). We have also performed electrical transport measurements on one more half-levitated sample (s3) and one more non-levitated sample (s4). The resistivity increases with decreasing temperature in $\rho$-$T$ curves of both samples (Fig.S3-4). Thus, all our samples show insulating behavior, no matter the samples are half-levitated (s2-3) or non-levitated (s1, s4). The fitted gap by band gap model is 207.29 meV for s3(Fig.S3b) and 204.31 meV for s4(Fig.S4b). The fitted gaps for all samples(s1-4) are closed to 200 meV. There is no significant difference for the gap value whether the sample shows half magnetic levitation or not.

Figure. 5c shows the *M-H* curves of s2 at 2 K and 300 K. The external magnetic field is swept from 40 to -40 kOe, and then from -40 to 40 kOe. The magnetization first increases quickly with increasing the magnetic field and then becomes saturation superposed with a linear diamagnetic behavior, similar with the non-levitated sample s1. Fig. 5d illustrates the *M-H* curves after subtracting the linear backgrounds, the saturated magnetization is ~$1.1 \times 10^{-2}$ emu/g at 300 K and ~$2.5 \times 10^{-2}$ emu/g at 2 K. As shown in Fig. 5e, the hysteresis loop on the *M-H* curves of s2 can maintain up to 300 K. We have also measured the ZFC and FC *M-T* curves from 2 K to 400 K with a field of 100 Oe. The measured *M-T* curves show similar branching behavior (Fig.5f) of FC and ZFC curves, consistent with results of s1. These results indicate the ferromagnetism with Curie temperature higher than 400 K in the half-levitated sample s2.

The magnetization measurements reveal the ferromagnetic order above room-temperature in both half-levitated and non-levitated samples. No Meissner effect is detected in our samples. The robust occurrence of ferromagnetism in a system composed of Pb, P, Cu, O is surprising. As suggested by theoretical calculations[13], Cu exists in a divalent form in LK-99 and may therefore carry a net spin magnetic moment. Thus, one possible mechanism for this ferromagnetism is the frustrated exchange interactions induced by the inhomogeneous distribution of the Cu substituents in the $Pb_{10-x}Cu_x(PO_4)_6O$ structure [22-23].

**Discussion**

The levitation behavior of LK-99 system in magnetic field has been widely reported [5-7, 18-19]. However, all the levitation behavior reported up to now is "half", since one side of the sample always touches the magnet. One must be very careful about the claim of superconductivity by this half-levitation under magnetic field since ferromagnetism could induce the same phenomenon [19]. Given the coexistence of diamagnetic and ferromagnetic signals in our samples and the samples from other groups [19-24], it is important to identify which signal dominates the half-levitation behavior. In our measurements, the contribution from diamagnetic quartz tube in our MPMS measurements is negligible (about -$10^{-7}$ emu at 300 K under magnetic field of 10 kOe). The diamagnetic susceptibility of the linear background is estimated to be ~-$2.7 \times 10^{-7}$ emu/gOe at 300 K for s1 and ~-$5.1 \times 10^{-7}$ emu/gOe for s2, respectively. This susceptibility is nearly in the same order of magnitude as copper (-$0.9 \times 10^{-7}$ emu/gOe). Thus, the impurities including Cu and its sulphides may contribute a significant value for diamagnetic background. Considering such a small magnitude, diamagnetism is less possible to dominate the half-levitation behavior. Furthermore, as illustrated in Extended Video.2, as the edge of the magnet approaches and moves away, sample s3 appears to have a tendency to point its geometric longest direction towards the direction magnetic induction line. The tendency means that the magnetization of s3 is along the longest axis. This is consistent with the scenario proposed before [19], in which ferro-magnetization prefers to align along the longest direction of the sample due to geometric anisotropy. Thus, we conclude that the half-levitation at least in our samples is due to ferromagnetism.

Then, the saturated ferromagnetic moments can be estimated by simply subtracting the linear backgrounds (Fig.3d and Fig.5d). For sample s1 without half magnetic levitation, the saturated ferromagnetic moments are ~0.6×10$^{-2}$ emu/g at 300 K and ~3.9×10$^{-2}$ emu/g at 2 K. For sample s2 with half magnetic levitation, the saturated ferromagnetic moments are ~1.1×10$^{-2}$ emu/g at 300 K and ~2.5×10$^{-2}$ emu/g at 2 K. Note that the ferromagnetic moments per unit mass for s2 are almost twice larger than s1 at room temperature, which may explain why s2 can be half-levitated while s1 cannot. Other relevant factors for the half magnetic levitation might include the center of gravity influenced by the mass distribution, susceptibility distribution, and so on. As an estimation, supposing the surface magnetic field of our Nd$_2$Fe$_{14}$B magnet to be 1 kOe and the mass center is the geometric center of our sample, we obtain the magnetic force moment of ~1×10$^{-9}$ Nm and the gravity moment of ~1.9×10$^{-9}$ Nm for s2. Since the magnetic force moment is in the same magnitude order of gravity moment, s2 can be half-levitated by the Nd$_2$Fe$_{14}$B magnet. On the other hand, the magnetic force moment is ~6×10$^{-10}$ Nm in s1, much smaller than the gravity moment ~2×10$^{-8}$ Nm of s1. Thus, it is reasonable that s1 does not show half-levitation behavior.

**Conclusions**

As a summary, we synthesized LK-99 like samples using solid-state sintering method and observed the half-levitation under the magnetic field of Nd$_2$Fe$_{14}$B magnet at room temperature. The structure and composition of the samples are characterized by XRD and EDS techniques. The results indicate that our samples are composed of inhomogeneous Pb$_{10-x}$Cu$_x$(PO$_4$)$_6$O, along with elemental Cu and Cu sulfides. The inhomogeneous distribution for different LK-99 samples may result in the distinct transport behaviors reported before [17,20-21]. The electric transport measurements on our samples reveal the band insulating behavior with a gap of ~200 meV. Furthermore, the magnetization measurements show the pronounced ferromagnetism superposed by a linear diamagnetic background, and the ferromagnetic hysteresis loops are maintained even at a temperature of 400 K. Our analyses suggest that the half magnetic levitation behavior is due to the ferromagnetism of the sample at room temperature. No superconducting signal, such as Meissner effect or zero-resistance, is detected in our samples. The ferromagnetism and an insulating gap of around 200 meV revealed by our systematic studies suggest LK-99 might be a promising ferromagnetic insulator or semiconductor.

**Availability of data and material:** The data and materials can be requested from corresponding author via e-mail.

**Competing interests:** All authors declare that there are no competing interests.

**Funding:** This work was financially supported by the National Natural Science Foundation of China (Grant No. 11888101) and the Innovation Program for Quantum Science and Technology (2021ZD0302403).

**Authors' contributions:** JW conceived and instructed the research. Under the guidance of JW, PW performed the magnetism measurements with the assistance of JG and SQ. PW, JG, YL and XL performed the transport measurements with the assistance of HJ and SQ. PW and XL analyzed the data under the guidance of JW. PW and CJ grew the single crystals. PW, JG, GM, CJ, SQ and JW wrote the manuscript with the input from all authors. YA conducted the XRD analysis.

**Acknowledgments:** We acknowledge the technical assistance from KG during sample growth.

# Figures

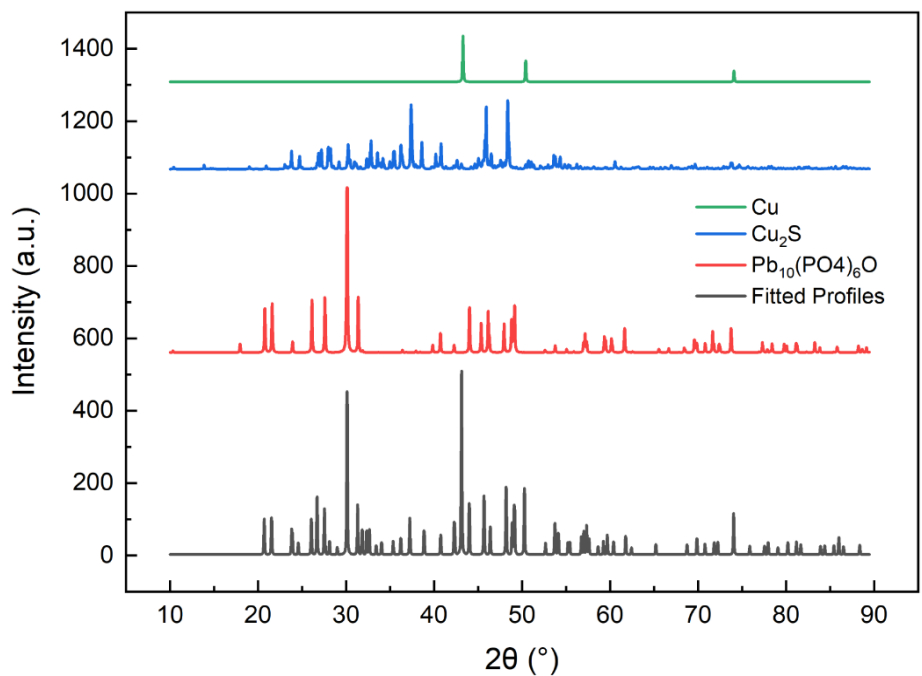

**Fig. 1.** XRD results of our powder samples at room temperature (black curve), together with standard XRD spectrum of $Pb_{10}(PO_4)_6O$ (red curve), $Cu_2S$ (blue curve) and Cu (green curve). The main XRD peaks correspond to $Pb_{10-x}Cu_x(PO_4)_6O$ and the other peaks represent the by-product, mainly including Cu and its sulphides.

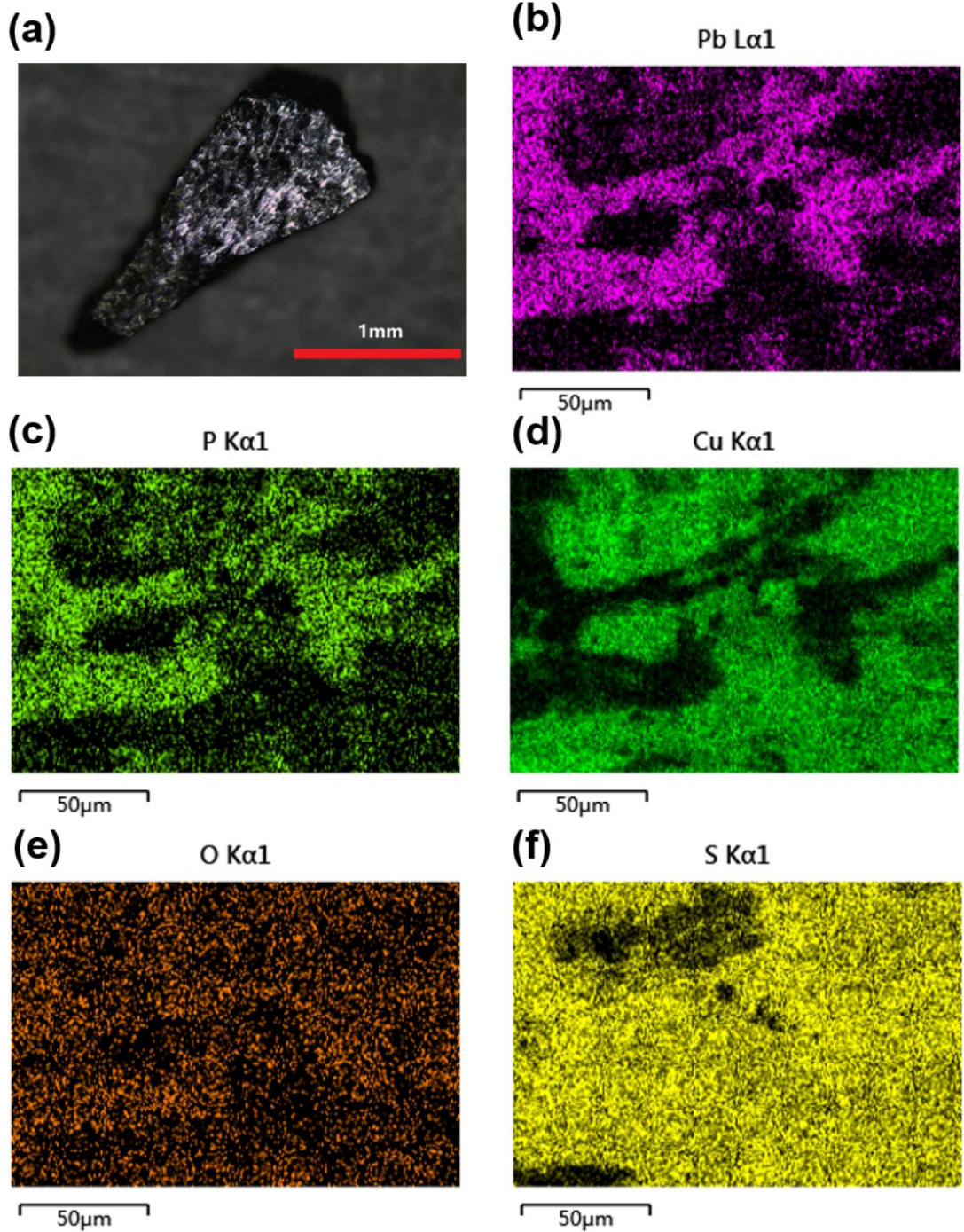

**Fig. 2.** (**a**) Optical image of non-levitation sample s1, the scale bar represents 1 mm. (**b-f**) Spatial mappings of energy-dispersive X-ray spectroscopy (EDS) results in s1, demonstrating the existence of Pb(**b**), P(**c**), Cu(**d**), O(**e**), and S(**f**). The distribution of Cu and Pb (or P) is nearly spatially complementary.

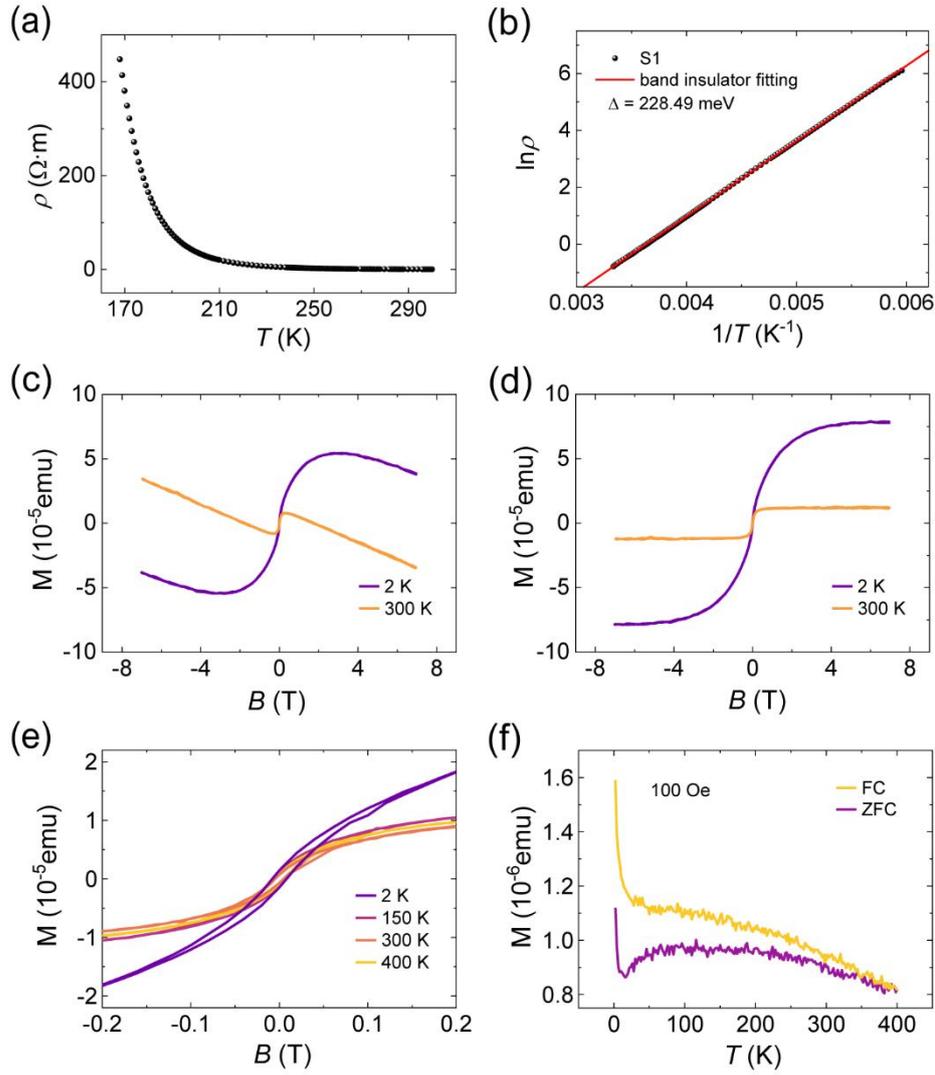

**Fig. 3.** Magnetization and electrical transport measurement results of sample s1. (**a**) Electrical transport properties of sample s1. Temperature dependence of resistivity ranging from 165 K to 300 K at zero magnetic field is shown. Typical insulating behavior is observed. (**b**) The fitting of temperature dependent resistivity with the band-gap model for insulators. The fitted gap is ~228.49 meV. (**c**) Magnetization versus magnetic field (*M-H*) curves at 2 K and 300 K. (**d**) *M-H* curves after subtracting linear backgrounds. (**e**) Zoom-in of *M-H* curves under nearly zero magnetic field. (**f**) zero-field-cooling (ZFC) and field-cooling (FC) *M-T* curves ranging from 2 K to 400 K under a magnetic field of 100 Oe.

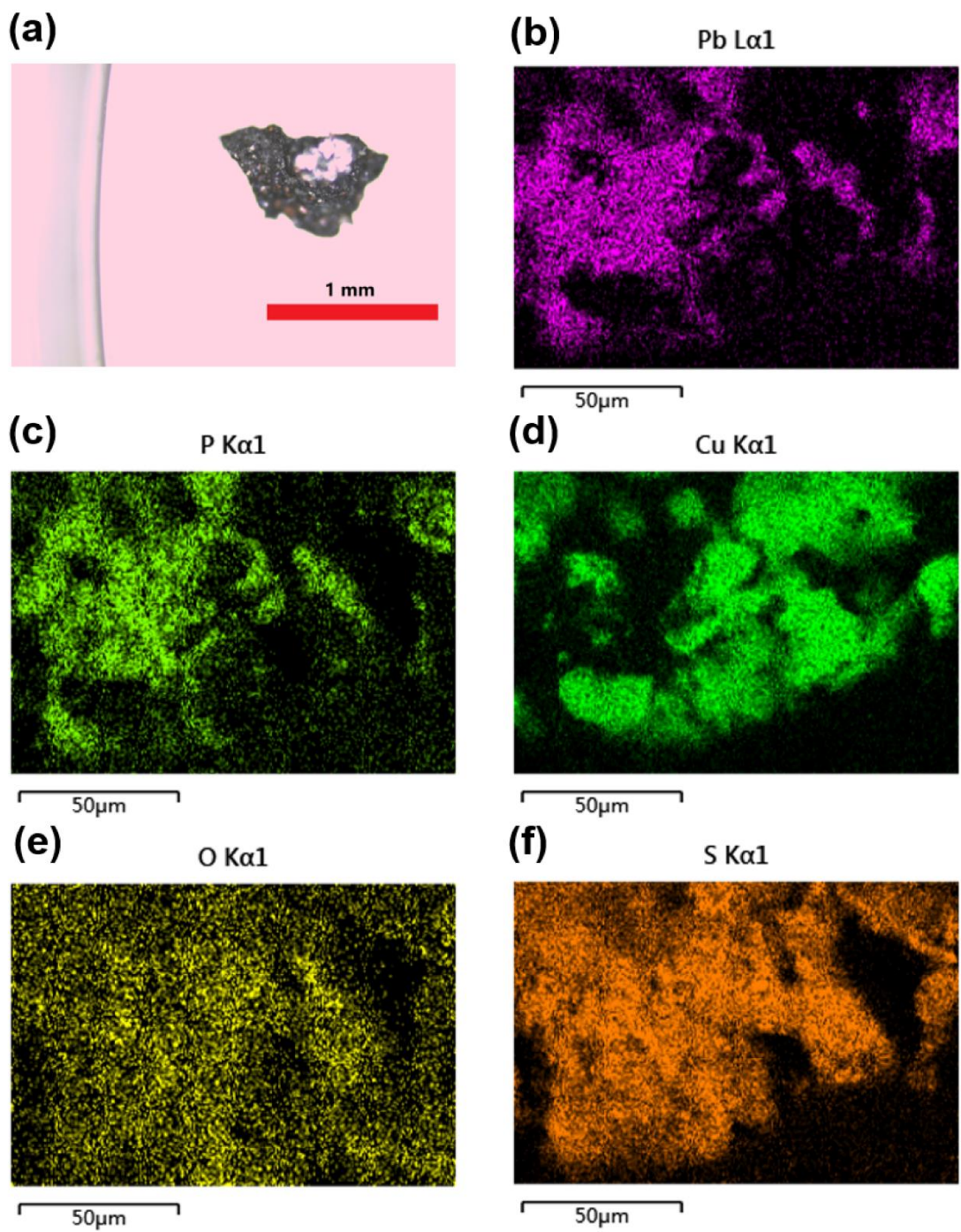

**Fig. 4.** (**a**) Optical image of sample s2 showing the half magnetic levitation at room temperature (see in Extended Video.1), the scale bar represents 1 mm. (**b-f**) Spatial mappings of energy-dispersive X-ray spectroscopy (EDS) results in s2, demonstrating the existence of Pb (**b**), P (**c**), Cu (**d**), O (**e**) and S (**f**). The distribution of Cu and Pb (or P) is nearly spatially complementary.

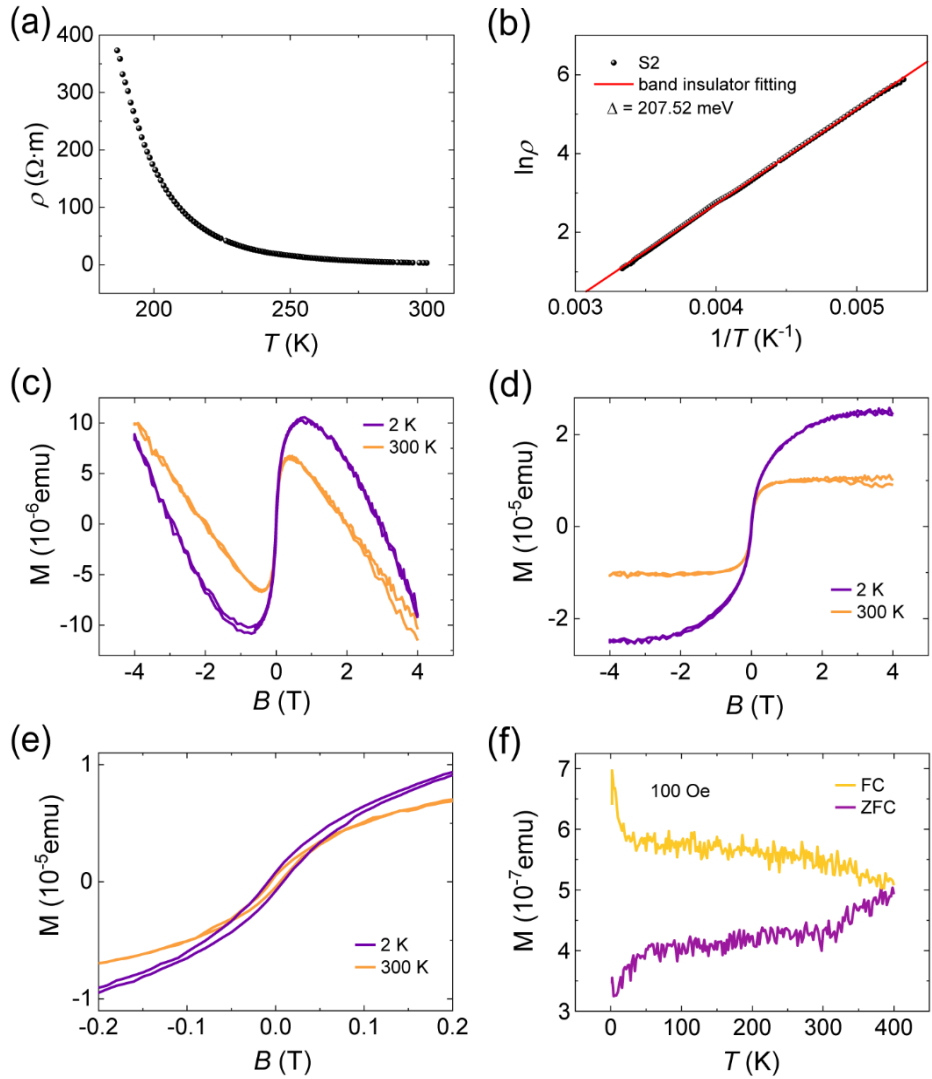

**Fig. 5.** Magnetization and electrical transport measurement results of sample s2. (**a**) Insulating behavior of the temperature dependence of resistivity ranging from 165 K to 300 K at zero magnetic field. (**b**) The fitting of temperature dependent resistivity with the band-gap model for insulators. The fitted gap is ~207.52 meV. (**c**) Magnetization versus magnetic field (*M-H*) curves at 2 K and 300 K. (**d**) *M-H* curves after subtracting linear backgrounds. (**e**) Zoom-in of *M-H* curves under nearly zero magnetic field. (**f**) Zero-field-cooling (ZFC) and field-cooling (FC) *M-T* curves ranging from 2K to 400 K under a magnetic field of 100 Oe.

# Supplementary materials

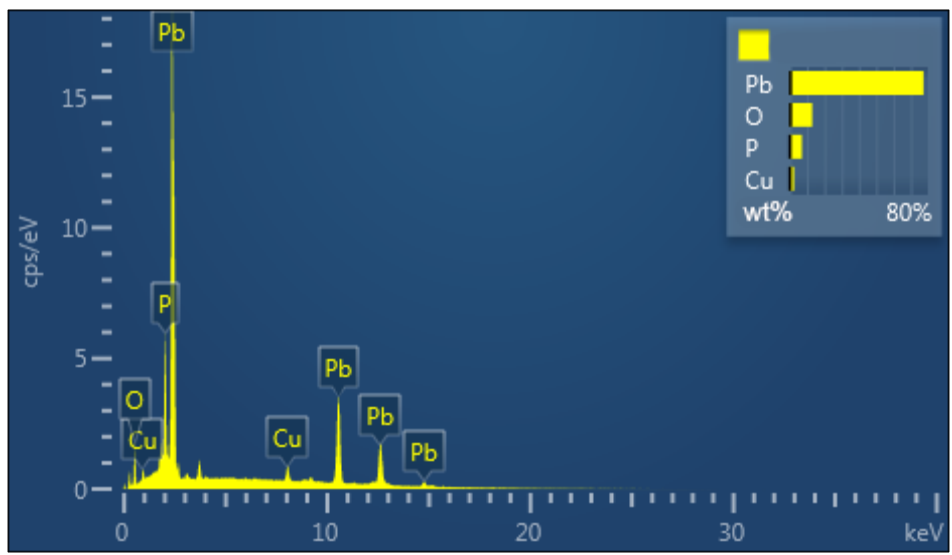

| Element | O | P | Cu | Pb | total |
|---|---|---|---|---|---|
| atomic percent | 55.83 | 15.7 | 2.67 | 25.8 | 100 |

**Fig. S1.** The EDS and the atomic percent of sample s1, indicating the presence of $Pb_{10-x}Cu_x(PO_4)_6O$.

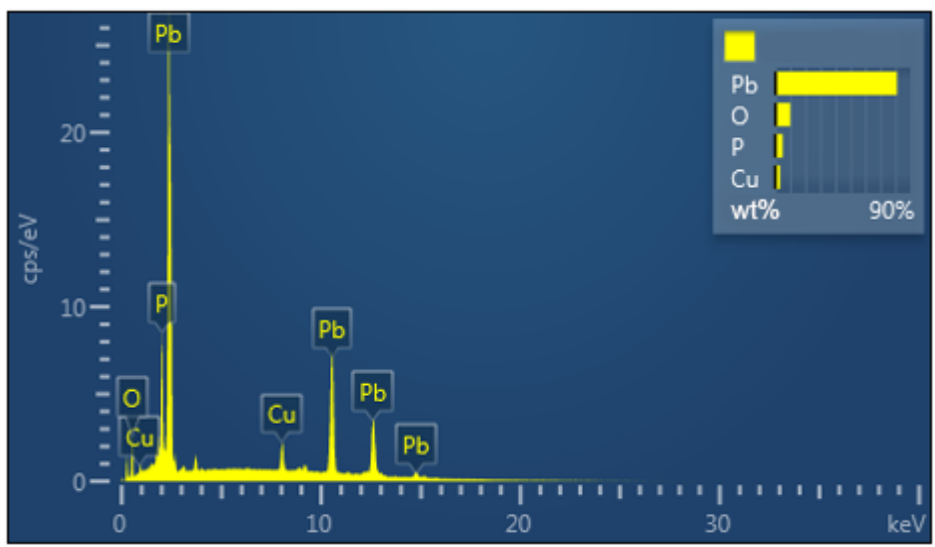

| Element | O | P | Cu | Pb | total |
|---|---|---|---|---|---|
| atomic percent | 50.96 | 13.34 | 4.43 | 31.27 | 100 |

**Fig. S2.** The EDS and the atomic percent of our sample s2, indicating the presence of $Pb_{10-x}Cu_x(PO_4)_6O$.

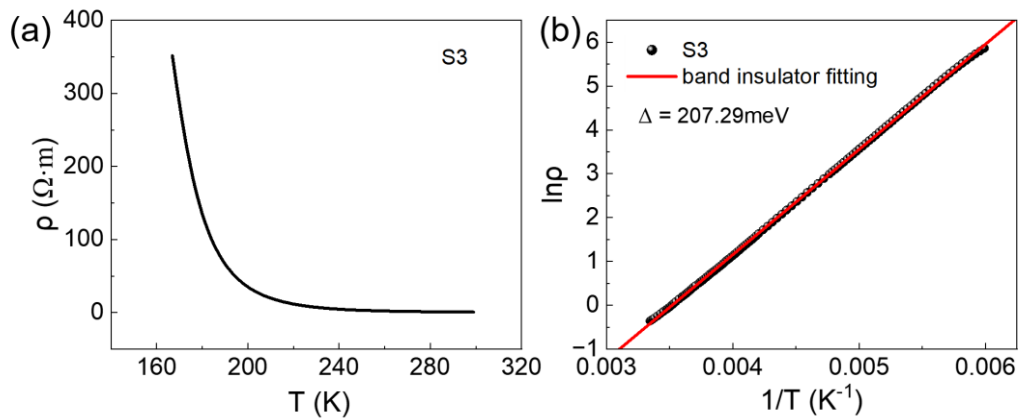

**Fig. S3.** Electrical transport measurement results of sample s3. (**a**) Insulating behavior of the temperature dependence of resistivity ranging from 165 K to 300 K at zero magnetic field. (**b**) The fitting of temperature dependent resistivity with the band-gap model for insulators. The fitted gap is 207.29 meV.

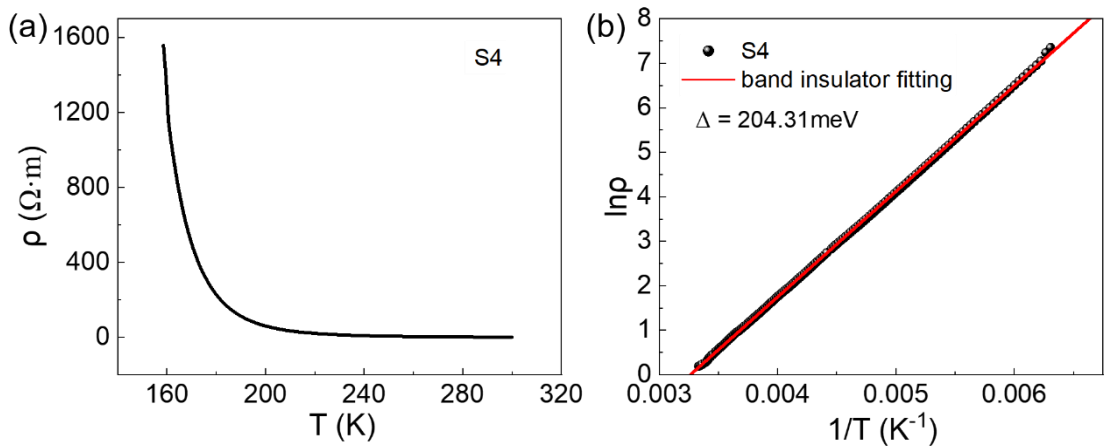

**Fig. S4.** Electrical transport measurement results of sample s4. (**a**) Insulating behavior of the temperature dependence of resistivity ranging from 165 K to 300 K at zero magnetic field. (**b**) The fitting of temperature dependent resistivity with the band-gap model for insulators. The fitted gap is 204.31 meV.

## Extended Videos

The snap shot of the half levitation behavior of sample s2 upon a $Nd_2Fe_{14}B$ magnet (Extended Video.1) is accessible on:
https://pku.instructuremedia.com/embed/65bc249e-dd55-4e50-80a4-062f366ae324
The snap shot of the half levitation behavior of sample s3 upon a $Nd_2Fe_{14}B$ magnet (Extended Video.2) is accessible on:
https://pku.instructuremedia.com/embed/2030dae3-61b3-47bd-a001-3a3b2eb9e210